# Fabrication of Nano-Gapped Single-Electron Transistors for Transport Studies of Individual Single-Molecule Magnets

J. J. Henderson, C. M. Ramsey, and E. del Barco[a]

*Department of Physics, University of Central Florida, 4000 Central Florida Blvd., Orlando, Florida 32816-2385*

A. Mishra and G. Christou

*Department of Chemistry, University of Florida, Gainesville, Florida 32611-7200*

a) Electronic mail: delbarco@physics.ucf.edu

**Abstract**

Three terminal single-electron transistor devices utilizing Al/Al$_2$O$_3$ gate electrodes were developed for the study of electron transport through individual single-molecule magnets. The devices were patterned via multiple layers of optical and electron beam lithography. Electromigration induced breaking of the nanowires reliably produces 1-3 nm gaps between which the SMM can be situated. Conductance through a single Mn$_{12}$(3-thiophenecarboxylate) displays the coulomb blockade effect with several excitations within ± 40 meV.

## I. INTRODUCTION

Electron transport properties of individual molecules have received considerable attention over the last several years due to the introduction of single-electron transistor (SET) devices[1-4], which allow the experimenter to probe electronic, vibrational[2] or magnetic[3,4] excitations in an individual molecule. In a three-terminal molecular SET the molecule is situated between the source and drain leads with an insulated gate electrode underneath. The insulating ligands on the periphery of the molecule act as isolating barriers, thus the couplings between the molecule and the electrodes are capacitive, and the magnitudes of which are primarily dependent on the molecule/lead distances and the ligand composition. Current can flow between the source and drain leads via a sequential tunneling process through the molecular charge levels, which the gate electrode is used to tune.

Conduction through a molecular SET only occurs when a molecular electronic level lies between the Fermi energies of the leads. A bias voltage, $V_{bias}$, applied between the source and the drain, changes the electrostatic potential of one of the leads by an energy $|eV|$. For small bias voltages, $|eV| < E_c + \Delta E$ where $E_c$ is the Coulomb charging energy and $\Delta E$ is the energy difference between consecutive charge states of the molecule being measured, current cannot flow though the device because the excited molecular levels are not available to conduct charges between the electrodes. This is known as the Coulomb blockade regime. A signature of this phenomenon is commonly seen at low temperatures as an absence of current for low bias voltages. As the bias voltage across the device increases, excited states will provide conduction channels in the device. As a result, discrete changes in the current through the SET will be obtained every time a new molecular level falls within the bias window. Applying a gate voltage moves these molecular states with respect to the electrodes Fermi levels.

Because the transport through single molecules proceeds via the discrete molecular levels,

one can "spectroscopically" measure the quantum energy landscape of an individual molecule rather than an ensemble of molecules. This is particularly important for the single-molecule magnets (SMMs), where it has been shown that the molecule's crystalline environment has a profound influence on the spin-Hamiltonian and therefore the quantum decoherence rates between $M_S$ levels.

SMMs are a class of molecules containing multiple transition metal ions bridged by organic ligands. These ions are strongly exchange coupled, often in a ferri-magnetic manner, to yield large magnetic moments per molecule. This large spin combined with zero-field splitting provides an anisotropy barrier to magnetization reversal. Of significant importance is the fact that quantum tunneling of the magnetization (QTM) can be observed between the different $M_S$ levels of the molecule. This behavior is illustrated by step-wise magnetic hysteresis loops and faster spin relaxation at certain resonance fields which switch on the quantum tunneling mechanism, a phenomenon known as resonant-QTM[5]. This unique feature of SMMs is a consequence of the quantum superposition of high-spin states of the molecule and has lead to the observations of a variety of fundamental phenomena, such as quantum (Berry-phase) interference between equivalent quantum tunneling trajectories[6]. Novel features of QTM are expected to manifest themselves in other observables as well. In particular, the effects of QTM on electronic transport remain to be explored in depth both experimentally and theoretically.

During the last year, there has been a significant effort in this direction by both experimental and theoretical groups. Van der Zant and coworkers reported Coulomb blockade and conduction excitations characteristic of a molecular SET in an individual $Mn_{12}$ SMM functionalized with thiol groups. Negative differential conductance and current suppression effects were explained in terms of relaxation between excited spin levels of the charged and uncharged states of the $Mn_{12}$[7]. Dan Ralph and coworkers reported transport through an individual $Mn_{12}$ SMM without any functionalization of the molecules. They found evidence of magnetic anisotropy in some of the studied molecules[8]. Even though these two results are very promising, they have failed to provide unambiguous evidence that transport occurred through an individual SMM that preserved key quantum properties found in their solid-state form (crystal). Several theoretical groups have proposed alternative ways to prove the SMM effect on the transport through a SET[9-13]. Leuenberger and Mucciolo showed that QTM interference effects (Berry phase) can be tuned to modulate the Kondo effect in an isolated SMM-based SET[10]. Similarly, Gonzalez and Leuenberger have recently shown that Berry phase effects can have a significant impact even in incoherent transport, away from the Kondo regime[13]. These results indicate that Berry phase interference might be the most promising way to characterize the unique transport properties of SMMs in comparison to conventional, nonmagnetic molecules.

In this context, we have successfully fabricated three-terminal SET devices and prepared thiophenecarboxylate functionalized $Mn_{12}$ derivatives, which bind to Au surfaces in order to begin probing the transport properties of an isolated SMM. The compound of interest in the present study is $Mn_{12}$(3-thiophenecarboxylate). This is a derivative of $Mn_{12}$-acetate for which the acetate group has been substituted with 3-thiophene carboxylate. The assembly of this molecule on Au and its bulk magnetic properties have previously been reported[14]. We have also synthesized the 2-thiophene carboxylate, 2-thiopheneacetate, and 3-thiopheneacetate variations for this study, but their results are not presented here.

In this report we present the technical aspects of our device fabrication and characterization as well as some preliminary results on transport through $Mn_{12}$ (3-thiophenecarboxylate).

**II. DEVICE FABRICATION**

Our SET devices are fabricated in a multi-step process on high-quality silicon substrates coated with a flat 1μm thermal oxide layer. The recipe we follow is closely adapted from that of reference 15. Our devices are fabricated using three layers of optical lithography and one layer of e-beam lithography. In the first two steps, the contacts and leads to which nanowires are attached are patterned followed by evaporation of Au with Cr for adhesion. The first is a thin 30 nm layer of gold, which protrudes just beyond the 200 nm thick second layer leaving a step to which the nanowire will overlap. The thick Au evaporation for the second layer works to minimize the device resistance and provide a good surface for wire bonding. Next, the gate electrode is patterned and Al is evaporated while cooling the substrate with liquid $N_2$. The gate is then allowed to oxidize overnight in atmospheric conditions, forming a 2-3 nm $Al_2O_3$ layer[16]. Finally, the nanowire pattern is written with electron beam lithography and gold deposited on that layer. An AFM micrograph of the completed device can be seen in Fig. 1 (right inset). At the narrowest part, the wire is 90 nm wide.

The nm size gap between the source and drain electrodes is formed by electromigration induced breaking of the nanowire[1]. Sourcing a voltage through the nanowire erodes away individual atomic components to the point where a very small gap forms with a large resistance that can be associated with the tunneling of electrons from source to drain. Fig. 1 shows the electromigration induced breaking ($T = 4$ K) of several wires fabricated following the process described above. The current through the nanowire is essentially Ohmic up to a voltage at which the nanowire breaks ($V_{break}$), producing a small gap. $V_{break}$ and thus, the gap size are tunable by adding series resistors[17] in line with the 30 $\Omega$ nanowire as the different sets of data in Fig. 1 illustrate.

The size of the gap created by electromigration is mainly determined by the current density flowing through the nanowire. For a given current density, the higher $V_{break}$ the larger the size of the gap created because stronger electric fields are present at the breaking point. In our devices, the post-breaking tunnel resistances vary between 10 k$\Omega$ and 100 G$\Omega$ for most of the wires. This huge variation corresponds to only a ~1nm variation in gap size according to previous studies[15]. Thus our gap sizes are generally in the 1-3 nm range. The critical current density necessary to break a gold nanowire is estimated to be $j_b = 5 \times 10^{12}$ A/m² [17] and the breaking current, $I_{break}$, necessary to achieve this characteristic current density depends on the cross sectional area of the nanowire. Therefore, $V_{break}$ is determined by the total resistance of the circuit as follows, $V_{break} = R_{Total} * I_{break}$, where $R_T = R_{nanowire} + R_{series}$. The histogram in Fig. 2 shows that the distribution of $I_{break}$, as measured from the electromigration data, is centered at 8.5 mA. Considering a cross sectional area of 90 nm (width) × 18 nm (thickness) for our nanowires, we obtain $j_b = 5.3 \times 10^{12}$ A/m², which is in excellent agreement with the accepted value[17]. If we associate the change in breaking currents ($\Delta I_{break} \sim 2$ mA) to variations of the nanowire thickness, this gives us a 3.4 nm thickness variation along the three-inch length of the silicon wafer used in fabrication. This variation is due to dispersion of the gold evaporation beam along the wafer (solid angle), since each point of the wafer is not equidistant from the source.

Fig. 3 shows four different characteristic I-V curves measured directly after electromigration induced breaking of the nanowire. The inset summarizes the percentage for which we observe each of the various curves. Here, we neglect to show the non-conducting curves, which occur about 50% of the time due to large (>3-4nm) gap formation. The IV characteristic curves of broken wires can then be grouped together as follows[18]: (a) *CB* curves with current suppression for low bias voltages consistent with Coulomb blockade effect; (b) *STP*, curves showing abrupt changes of current (steps) consistent with Coulomb blockade effect in molecular SETs; (c) *ZBE*, zero bias enhancement of the conductance consistent with low temperature Kondo effect; and (d) *SMH*, smooth asymmetric curves not crossing $I=V=0$. The curves shown occur on different current scales ranging from µA down to tens of pA but are normalized to µA for ease in comparison of their features.

**III. SINGLE-MOLECULE TRANSPORT MEASUREMENTS**

The Mn$_{12}$(3-thiophenecarboxylayte) for SET studies was prepared by dissolving a polycrystalline sample in CH$_2$Cl$_2$ (0.1 mM). The devices were cleaned by oxygen plasma and immediately immersed into the solution for about 90 minutes. After the allotted time, the sample was removed from the solution, rinsed with CH$_2$Cl$_2$ and blown dry with a stream of N$_2$. Following self-assembly of the molecules, electromigration-breaking and measurements were performed at 4 K.

Fig. 4 displays the differential conductance (*dI/dV*) contour-plotted as a function of the bias and gate voltages (the added lines are to help display the linear conductance excitations present in this particular measurement). The presence of multiple parallel excitations forming the characteristic diamond shape of a molecular SET indicate the crossing between two charge states (*N* and *N*+1) of the molecule, and reveal the complex nature of the charge states of this particular molecule. The excitations are found in an energy range (0 ± 40 meV) similar to what has been observed by other authors in Mn$_{12}$-based SETs[7,8]. We have observed several other molecules behaving similarly in experiments carried out at 4 K in the same set of devices; however many of the molecules measured were not stable enough to measure for long periods of time. Out of 60 electromigration-broken wires, we have seen 3 that were stable enough to measure.

**IV. CONCLUSIONS**

We have outlined the technical details of our single electron transistor devices and demonstrated that they can be used to measure single-electron transport through individual Mn$_{12}$ SMMs. Our current objective is to probe different SMMs at low temperatures (>15mK) and in the presence of high magnetic fields generated by a vector superconducting magnet which allows arbitrary orientation of the magnetic field with respect to the geometry of the SMM-based SET.

**V. ACKNOWLEDGEMENTS**

J.J.H. thanks the I$^2$Lab at UCF for financial support.

**FIGURE CAPTIONS**

**Figure 1.** *IV* curves recorded during electro-migration-induced wire breaking for three series resistances; (left inset) Schematic representation of a three terminal SET; (right inset) AFM image of a SET device.

**Figure 2.** Distribution of breaking currents for a series resistance of 225 Ω. The distribution is associated to changes in the thickness, *t*, of the gold nanowire.

**Figure 3.** Selection of different *IV* curves observed after wire breaking: asymmetric smooth curves (SMH), curves with zero bias enhancement of the current (ZBE), curves with steps (STP) and curves with current suppression at low voltages consistent with Coulomb blockade (CB); (Inset) pie chart showing statistics of the *IV* curve types found after the breaking of over 60 nanowires.

**Figure 4.** Conductance (*dI/dV*) contour plot as a function of the bias and gate voltages, showing the typical Coulomb diamond response of a molecular SET. Interestingly, at least seven excited states are observed.

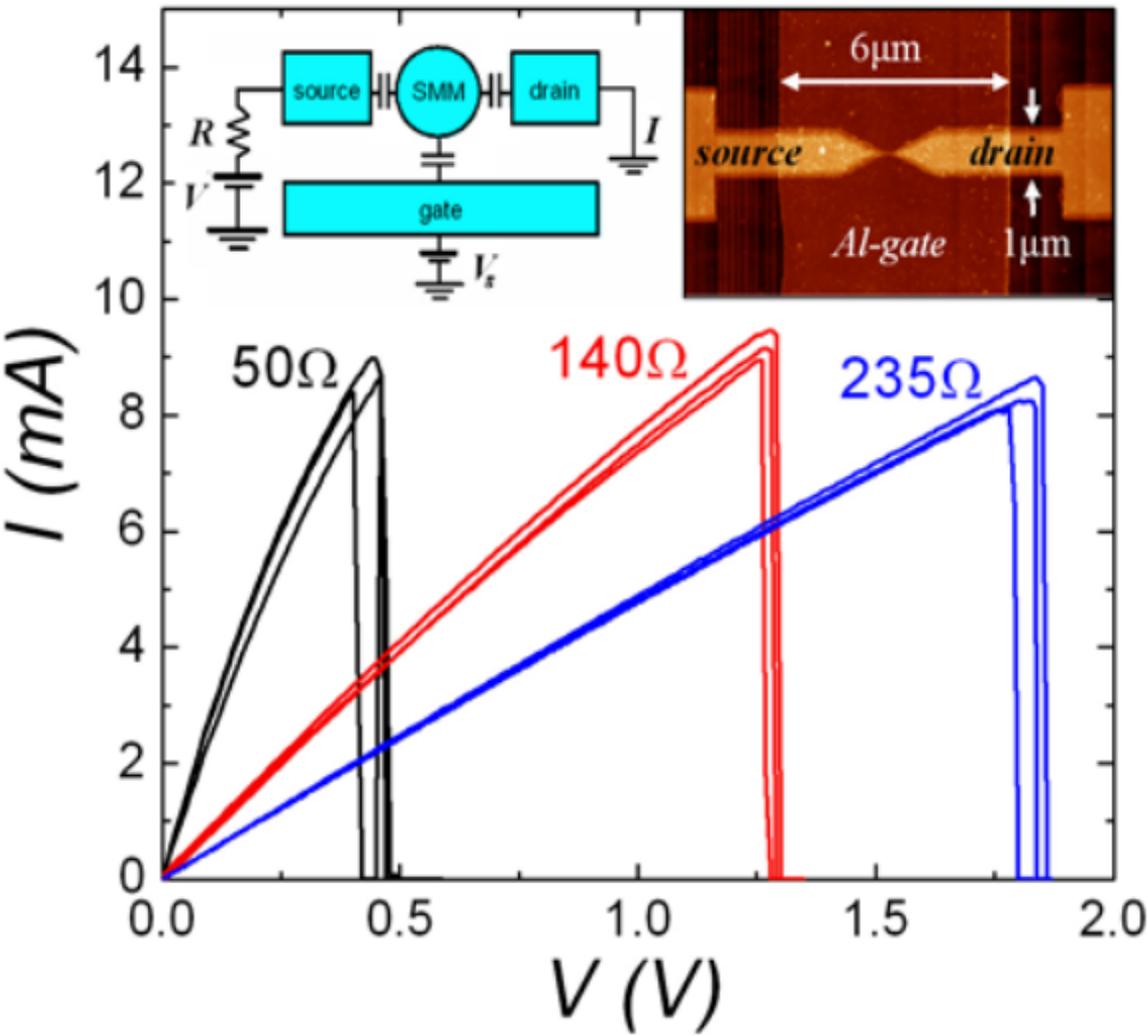

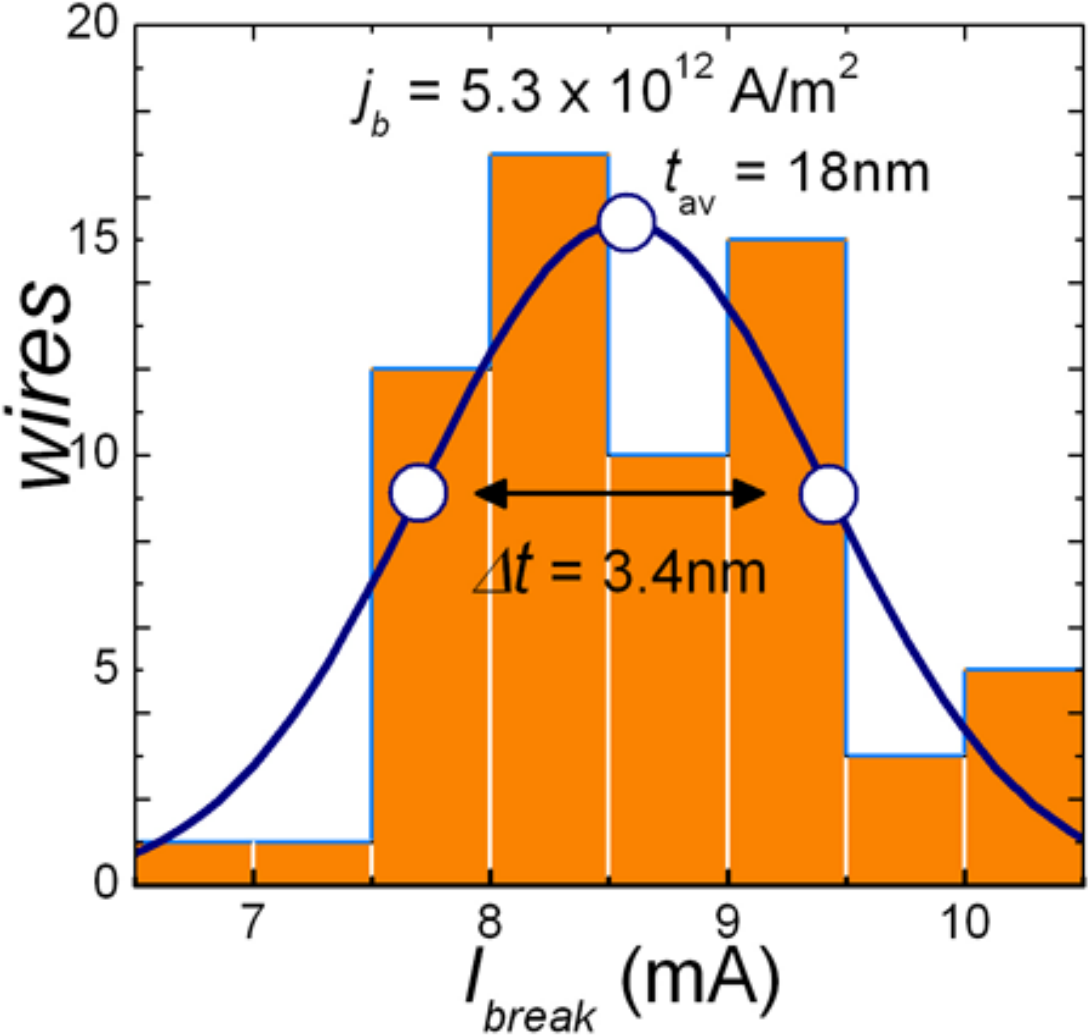

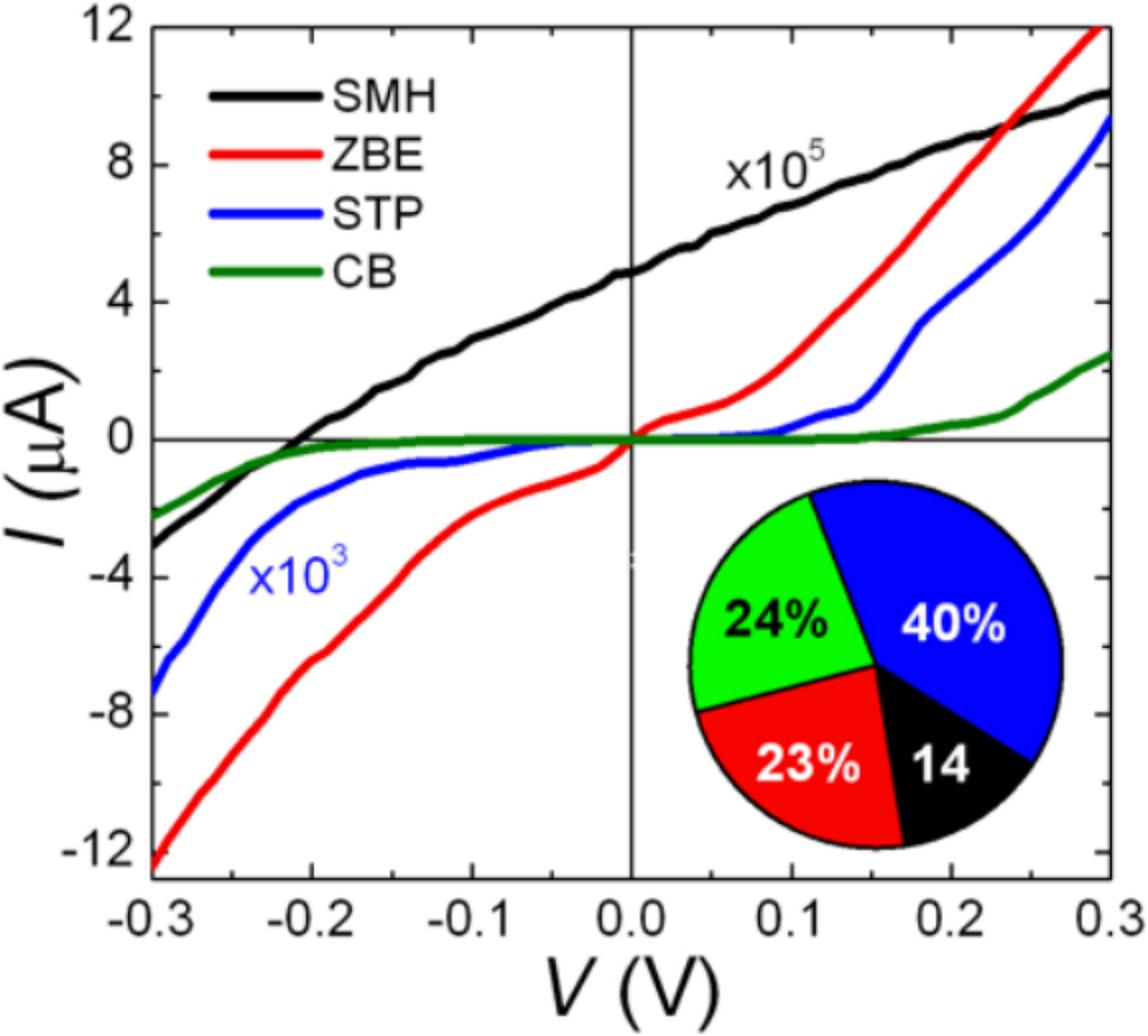

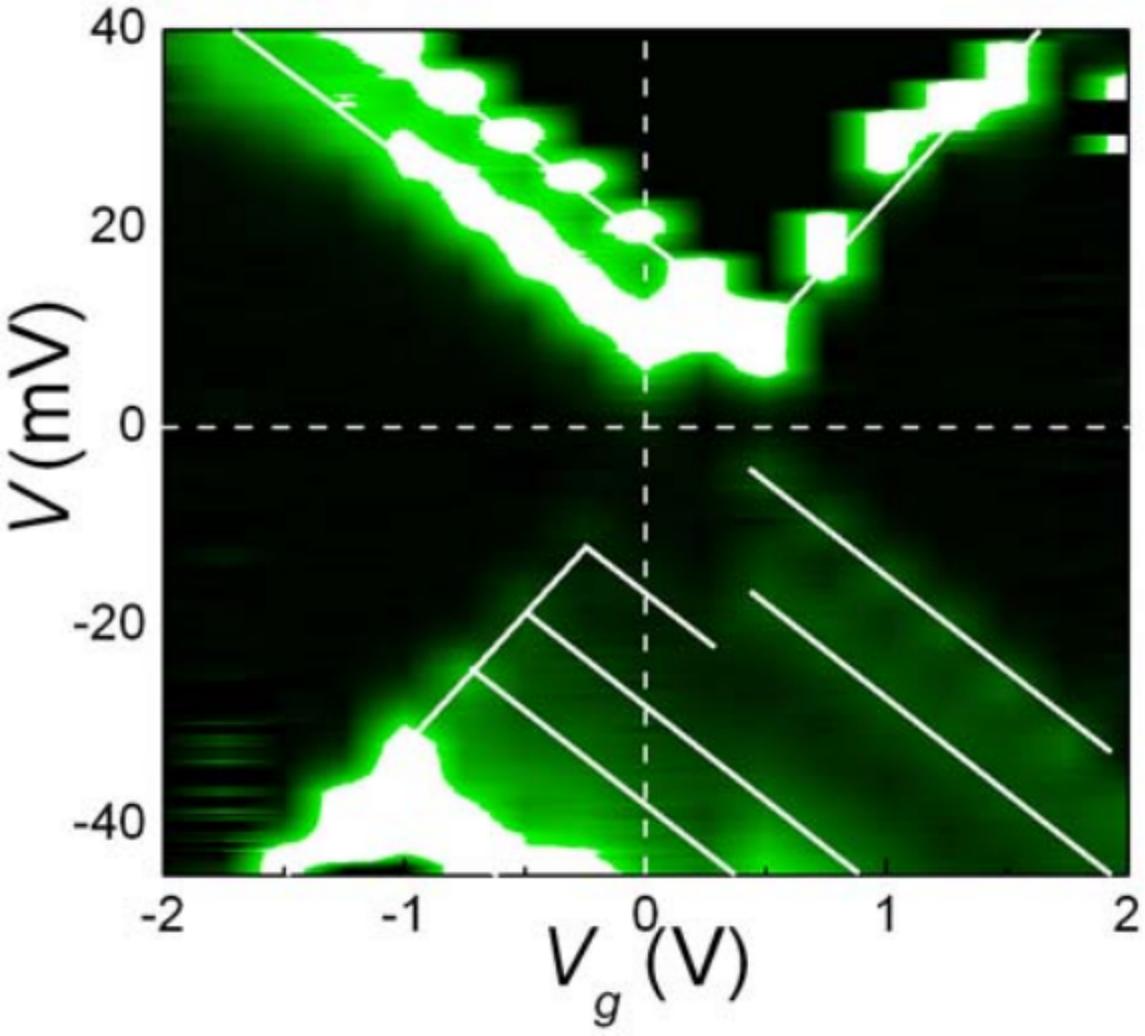